\documentclass[12pt]{article}

\usepackage{epsfig}
\usepackage{graphicx}
\usepackage{amssymb,amsmath}

\usepackage{epsfig,cite}
\newcommand{\beq}{\begin{equation}}
\newcommand{\eeq}{\end{equation}}
\newcommand{\be}{\begin{eqnarray}}
 \newcommand{\ee}{\end{eqnarray}}

\newcommand{\ov } {\over }
\newcommand{\p }{\partial }

\def\appendix#1{
  \addtocounter{section}{1}
  \setcounter{equation}{0}
  \renewcommand{\thesection}{\Alph{section}}
  \section*{Appendix \thesection\protect\indent \parbox[t]{11.15cm}
  {#1} }
  \addcontentsline{toc}{section}{Appendix \thesection\ \ \ #1}
  }

\begin{document}
\null\vskip-24pt
\hfill {\tt hep-th/0602125}
\vskip2truecm
\begin{center}
\vskip 0.2truecm {\Large{\bf Massless radiation from Strings:} \\ 
{\it quantum spectrum average statistics and cusp-kink configurations}}
\vskip 0.2truecm

\vskip 0.7truecm
\vskip 0.7truecm

{\bf Roberto Iengo}\\
\vskip 0.4truecm
\vskip 0.4truecm

{\it  International School for Advanced Studies (SISSA)\\
Via Beirut 2-4, I-34013 Trieste, Italy} \\
{\it  INFN, Sezione di Trieste}

\medskip

\end{center}
\vskip 2truecm 

\noindent\centerline{\bf Abstract}  
\vskip0.5cm
We derive general formulae for computing the average spectrum for Bosonic or Fermionic massless emission from generic or particular sets of closed superstring quantum states, among the many  occurring
at a given large value of the number operator. In particular we look for states 
 that can produce a Bosonic spectrum resembling
the classical spectrum expected for peculiar cusp-like or kink-like classical configurations,
and we perform a statistical counting of their average number. The results can be relevant 
in the framework of possible observations of the radiation emitted by cosmic strings.

\newpage
\section{Introduction}
In this paper we study the quantum massless radiation, both Bosonic and Fermionic,
from excited closed superstrings.
(For a general study of the decay of superstrings by precise numerical computations
see \cite{Okada, CIR1, CI, CIR2, HAND} ).
 In particular we look for string states that, in some range of  the radiation energy,
produce a spectrum
with some (bosonic) characteristics found in the classical approximation, namely the
interesting  cases of classical {\it cusps} or {\it kinks} \cite{book,Damour}.

The quantum spectrum is expected to agree possibly with the classical rersults
in a low energy range, that is for wavelengths much larger than $\sqrt{\alpha'}$, of course.

\vskip0.2cm

We will try to be as general as possible and therefore we do not  write in detail 
particular string states that  give the a particular spectrum. 
Rather, we find that a spectrum resembling, for instance, classical cusp-kink characteristics
occurs on average for string configurations in which the mode excitations 
satisfy a kind of sum rule.  We can thus further count the number of   such strings 
satisfying that constraint.
This result  can be useful for evaluate how much it is likely to find that particular spectrum
among the various signals possibly arriving from cosmic strings (for general studies of cosmic 
strings see \cite{witten, book, myers, kibble, Manes, Manes2, Manes3} ).
\vskip0.2cm

Our method is based on the observation that average radiation spectra from strings 
and their properties are easily derived in a suitable LightCone (LC) gauge, thus working
directly with physical states and avoiding the ghost formalism.
\vskip0.2cm

The Section 2 summarizes (and generalizes) the classical analysis.  
\vskip0.2cm

In Section 3 we introduce the convenient LC gauge and we
derive the main quantum formula. It is surprisingly simple both for Bosonic and
Fermionic massless emission.

\vskip0.2cm

In Section 4 we derive the generic (Bosonic and Fermionic)
average radiation spectrum from a string of large mass, 
by using the quantum formula in the  LC gauge and the standard statistical mechanics
method of the chemical potential.
\vskip0.2cm

In Section 5 we modify the chemical potential by
introducing a suitable constraint in the average, in the form of a weight
depending on a sum over the mode occupation numbers. 
We show that, in this way, a {\it cusp} or {\it kink} like spectrum is obtained 
in some sizable radiation energy domain. We then estimate the number of such string states
and how rare they are among the generic set of states of a given large mass.
\vskip0.2cm

In the Appendix A we review the (classical) relation between the gauge where 
$\p X^0$ is constant
(which we call the TP Temporal gauge) and the gauge where $\p X^+$ is constant
(which is called the LC LightCone gauge). That relation can also be seen as an algorithm
for obtaining (classical) solutions in the TP gauge which will automatically satisfy 
the Virasoro constraints.  
 
\vskip0.2cm

In the Appendix B we construct a sample of a generic classical string state in the Temporal gauge, following the results of Sect.4 and the recipes of Appendix A (for a recent portait of the string based on the computation of form factors see \cite{Manes3}).   

\vskip0.2cm

In the Appendix C we discuss the particular case of the state of maximal angular momentum, which
classically is an example of a cusp. We compute and compare the classical and the quantum spectrum 
for the mass $M=\sqrt{4 N/\alpha'}$ with $N=1000$. Even for this large value of $N$, there is no 
radiation energy region for which the quantum spectrum matches the classical cusp behaviour. The classical and quantum spectra show a good agreement for 
small radiation energy, where most of the radiation occurs, but
where however the classical behaviour is not yet of the cusp form;
for larger radiation energies, where the classical spectrum  is cusp-like, the quantum
spectrum falls off to zero much more rapidly than the classical one.

\section{The classical computation}
We begin by reviewing the classical massless radiation rate of a closed string (see \cite{Damour},
\cite{CIR2} ):
\beq
rate= g_s^2 {p_0^{D-3}\ov M^2}\int d\Omega \sum_{\xi,\tilde\xi}|\xi_j I_R^j \tilde\xi_k I_L^k|^2
\label{rate}
\eeq
Here $D$ is the number of extended dimensions,
$M=\sqrt{4N/\alpha'}$ is the mass of the string (assumed at rest), $p_0$ is the energy of the emitted massless
particle (graviton, scalar or antisimmetric tensor), $\xi_j\tilde\xi_k$ its polarization 
and
\beq
I_{R,L}^j=\int d\sigma_{\pm} e^{ip_{\mu}X^{\mu}_{L,R}(\sigma_{\pm} )} \p X^j_{R,L}(\sigma_{\pm} )
\eeq 
This computation is usually done in the {\it Temporal gauge} where 
$X^{0}_{L,R}=\alpha' (M/2) \sigma_{\pm}$ with \\
$\sigma_{\pm}=\tau\pm\sigma$ such that  $X^0=X^0_L+X^0_R=\alpha' M\tau$.
 
\vskip0,2cm

There is a saddle point \cite{Damour} in the integral defining $\xi\cdot I_{L,R}$  if 
\beq
p\cdot\p X_{L,R} =0 ~at~some~\sigma_{\pm}=\sigma^c_{\pm}
\label{cusp1}
\eeq
We can take $\sigma^c_{\pm}=0$. This is the condition defining a $ {cusp}$.  \\
Another interesting case is when $p\cdot\p X_{L,R}$ has a discontinuity.
This is referred as the ${kink}$.  

To be precise,  a $cusp$ or a $kink$ occur when the above conditions are respectively satisfied 
in both the Left and the Right sectors. 

However, since our study can be done separately and independently for each sector, {\it from now on we  discuss, say, the Left sector only and write $X$ meaning $X_L$ and $~\xi\cdot I$ meaning $\xi\cdot I_L$
and $\sigma$ meaning $\sigma_{+}$.} Of course the other ($R$) sector is treated in the same way.
\vskip0.2cm

Take the frame where $p_{\mu}=(p_0,p_z,0)$ with $p_z= -p_0$. 
In this frame $p_{+}={p_0+p_z\ov \sqrt{2}}=0$ and the cusp condition
is $\p X^{+}=0$.

In the "temporal gauge"  the $cusp$ is only possible
if also $\p X^T=0$ (for every $T$-ransverse component). 
In fact in the Temporal gauge $\p X^{+}+\p X^{-}=~constant$ and
it follows from the classical Virasoro constraints  $2\p X^{+}\p X^{-}=(\p X^T)^2$
that if $\p X^T\to 0$  then $ \p X^{+}\sim (\p X^{T})^2$.

Assuming that $\p X^T$ vanishes linearly  we have
\beq
p_{-} X^{+}\sim \sigma^3
\label{cusp}
\eeq
In this case, for large $N_0\equiv \sqrt{\alpha'N}p_0$ we can extend the integration over $\sigma$
to $-\infty ,+\infty$ and we get
\beq
\xi\cdot I=\int d\sigma e^{ip_{-}X^{+}} \xi\cdot \p X\sim \int d\sigma \sigma e^{icN_0\sigma^3}
\sim N_0^{-2/3}
\eeq 

In general it could be that $\p X^T\sim \sigma^{\beta}$ and thus 
$p_{-} X^{+}\sim \sigma^{2\beta+1}$.  In this case 
\beq
\xi\cdot I\sim N_0^{-{\gamma\ov 2}}~ ~with~~1< {\gamma}={2\beta+2\ov 2\beta+1}\leq 2
\label{general}
\eeq

This general behavior in $N_0$ includes the result for the $kink$ for which 
$\beta =0\rightarrow\gamma =2$. 

{\it We will refer to all these cases with the various possible $\gamma$ as "cusp"}.

\section{The quantum computation}

The quantum expression of the rate is the same as eq.(\ref{rate}) with $\xi\cdot I$ 
given by the relevant quantum matrix element \cite{CIR2}.
The quantum computation is most easily done in a Light-Cone (LC) gauge, where 
the Fock space of the $T$-ransverse oscillators comprises all the physical states.
We specify the LC gauge by taking LC coordinates such that $p_{+}=0$ as
above. \\
In the LC-gauge  $X^{+}=\alpha' (M/2) \sigma$ (remember that we mean 
$X^+_L=\alpha' (M/2) (\sigma+\tau)$).

\vskip0.2cm

{\it Also  the classical computation for the cusp can be performed in the LC-gauge.}

This amounts simlpy to a change of integration variable in eq.(\ref{cusp})
\beq
\sigma\to\sigma'=\sigma^{3}
\label{classicalLC}
\eeq 
In this gauge, that is in this new variable, $\p X^T$ is divergent 
$\sim \sigma'^{-1/3}$ (or $\sim \sigma'^{-\beta/(2\beta+1)}$
in the general case), rather than going to zero (see also the Appendix).

\vskip0.2cm

{\it {Quantum computation.}}
In the $p_+=0$ LC gauge the (Left or Right) part of the vertex operator for emitting
a massless NSNS state is (we can take it at $\sigma=0$)
\beq
V(0) = \xi_T\cdot\p X^T(0) e^{ip_{-}X^{+}(0)}=\xi_T\cdot\p X^T(0)
\label{vertex}
\eeq

Thus we have
\beq
|\xi\cdot I|^2=\sum_f |<f|\xi_T\cdot\p X^T|\Phi_N>|^2
\eeq
where $|\Phi_N>$ represents the radiating state with mass $M=2\sqrt{N/\alpha'}$,
which is supposed to be at rest,
and $|f>$ is a possible final state with mass $M'=2\sqrt{N'/\alpha'}$.
Let us take here $\alpha' =4$.
The radiated energy is $p_0 =(M^2 -M'^2)/2M=N_0/2\sqrt{N}$ with $N_0=N-N'$.

The vertex is linear in the transverse oscillator operators and the computation  is easy.

\vskip0.2cm

We are interested in the average (in particular  for $N\to\infty$) over the many different states $|\Phi_N>$
which share some properties. 

\vskip0.2cm

For a definite
$p_0=N_0/2\sqrt{N}$, that is for a definite $N_0$, we have
\beq
<|\xi\cdot I_{L}(p_0)|^2>= {1\ov {\cal N}} Tr[ (\xi_T\cdot\p X^T)^\dagger_{N_0}~ 
(\xi_T\cdot\p X^T)_{N_0} ]_N
\eeq
The trace is restricted to initial states with fixed
$N$ (${\cal N}$ being their number) and moreover
$(\xi_T\cdot\p X^T)_{N_0}$ means restricting the operator to that part that lowers 
the value of  the number operator 
$\hat N$ from $N$ to  the final state value $N'=N-N_0$. 

In terms of the transverse oscillators
\beq
 (\xi_T\cdot\p X^T)_{N_0}=\sqrt{\alpha'\ov 2}N_0^{1/2}\xi\cdot a_{N_0}
\eeq
The  normalization is $[a^i_{+n},a^j_{-m}]=\delta^{ij}\delta_{nm}$ and we take $\alpha' =4$.

\vskip0.2cm

Therefore, $p_0=N_0/2\sqrt{N}$ being the radiated energy, we get  THE MAIN FORMULA
\beq
\sum_{\xi}<|\xi\cdot I(p_0)|^2> =\sum_{\xi}<[V^\dagger V]_{N_0}> =
\sum_{\xi}<|\xi\cdot\p X|^2_{N_0}> = 2<N_0a_{-N_0}\cdot a_{N_0}>
\label{main}
\eeq
 
The above formula describes the NS radiation (say, in the Left sector). 
In the LC one easily get also 
the corresponding formula for the R(amond) radiation by using the Green-Schwarz formalism
(see \cite{GSW} ). \\
We remember that in this formalism the fermionic degrees of freedom are carried by 
the $S^a_n$ oscillator operators, $a=1,\cdots ,8$ being a spinor index, satisfying
$\{S^a_{+n},S^b_{-m}\}=\delta^{ab}\delta_{nm}$. \\
Since the emitted momentum satisfies $\vec p_T=p_+=0$, the vertex for emitting the (Left part of)
a massless fermion is $V_F(0)=u\cdot S(0)\sqrt{ \hat P_+}$
where $u(p)$ is a suitably normalized polarization spinor and $\hat P_\mu$ is the momentum operator. \\
By averaging over $u$ (we take $\sum_u u_a^\dagger u_b=2\sqrt{\alpha'}p_0\delta_{a,b}$) we find
\beq
\sum_{u}<|u\cdot I_F(p_0)|^2> =\sum_{u}<[V^\dagger_F V_F]_{N_0}^2> =
2<N_0S_{-N_0}\cdot S_{+N_0}>
\eeq
We remember that in terms of the LC oscillators the number operator is \\
$\hat N=\sum_{n>0}(na_{-n}\cdot a_{+n}+nS_{-n}\cdot S_{n})$.
\vskip1cm

\section{The average spectrum}
{\it We will call} $ <2na_{-n}\cdot a_{n}>$ {\it the spectrum for the Bosonic radiation, 
or $ <2nS_{-n}\cdot S_{+n}>$ for the Fermionic one,
although this is only the Left part, and 
to obtain the physical spectrum one has to take the product of Left and Right times 
the phase space $ {p_0^{D-3}\ov M^2}\Omega$.}
\vskip0.5cm

{\it {Now we review the derivation of the general average spectrum}}  
that is taking the average over all the states with $<\hat N>=N$ \cite{Amati2}, \cite{CIR2}. 

Mimicking statistical mechanics introduce a chemical potential term $e^{-\hat N\epsilon}$ 
and, beginning with the Bosonic spectrum, replace
\beq
Tr[na_{-n}\cdot a_{n}  ]_N\to Tr[na_{-n}\cdot a_{n} e^{-\hat N\epsilon} ]
\eeq
 and we fix $\epsilon$ requiring $<\hat N>=N$.
 We get
\be
<na_{-n}\cdot a_{n}>+<nS_{-n}\cdot S_{n}>=
 {1\ov {\cal N}}Tr[(na_{-n}\cdot a_{n} +nS_{-n}\cdot S_{n})e^{-\hat N\epsilon} ]= \\ \nonumber
= D_T~n~\{ {e^{-n\epsilon}\ov 1-e^{-n\epsilon}}+{e^{-n\epsilon}\ov 1+e^{-n\epsilon}}\}
\ee
where $D_T$ is the number of tranverse dimensions.

Similarly
\beq
{\cal N}=Tr[ e^{-\hat N\epsilon} ] =\prod_n ( {1+e^{-n\epsilon}\ov 1-e^{-n\epsilon}} ) ^{D_T}
\eeq
$\epsilon$ is fixed by requiring
\beq
N=D_T\sum_n\{ {ne^{-n\epsilon}\ov 1-e^{-n\epsilon}}+{ne^{-n\epsilon}\ov 1+e^{-n\epsilon}}\}=
-{d\ov d\epsilon}D_T\sum_n  \log[{1+e^{-n\epsilon}\ov 1-e^{-n\epsilon}}]
\eeq
(we recognize the standard saddle-point equation of string theory).

For small $\epsilon$
\beq
D_T\sum_n \log [{1+e^{-n\epsilon}\ov 1-e^{-n\epsilon}}]
\to {D_T\ov \epsilon}(c_F+c_B)
\eeq
with $c_F= \int dx \log [1+e^{-x}]=\pi^2/12$ and $c_B= - \int dx \log [1-e^{-x}]=\pi^2/6$. 

\vskip0.2cm

Therefore 
$\epsilon = \sqrt{D_T(c_F+c_B)/N}$ and we get for large $N$
\be
\log[{\cal N}]\sim {2\sqrt{ND_T(c_F+c_B)}} \\ \nonumber
\sum_{\xi}|\xi\cdot I(p_0)|^2 = 2{ N_0 e^{-N_0\sqrt{D_T(c_F+c_B)/N}}\ov 1-e^{-N_0\sqrt{D_T(c_F+c_B)/N}}}
\label{spectrum}
\ee
(remember $N_0=\sqrt{\alpha'N}p_0$). Thus we get a thermal-like ($Left$ part of the) spectrum with a temperature $\sim 1/\sqrt{\alpha'}$.

By repeating the computation for the ($Left$ part of the) Fermionic spectrum it easily seen that one gets 
a Fermi-Dirac distribution 
\beq
\sum_{u}|u\cdot I_{F}(p_0)|^2\sim{ N_0 e^{-N_0\sqrt{D_T(c_F+c_B)/N}}\ov 1+e^{-N_0\sqrt{D_T(c_F+c_B)/N}}}
\label{Fspectrum}
\eeq
\vskip1cm

\section{The average quantum cusp-kink-like spectrum }
{\it { Looking for the quantum states corresponding to the classical $cusp$ or $kink$.}} 
In this case we look for the Bosonic radiation.

The classical {\it cusp} expression for $\xi\cdot I$  eq.(\ref{general}) is obtained for a particular 
angle of the radiation momentum $\vec p$, namely the one for which $\p X^+(\sigma)=0$ 
at some $\sigma$, where $X^+= (p_0X^0-\vec p\vec X)/p_0$. Taking the $p_+=0$ Light-Cone 
frame and looking for a quantum spectrum corresponding to the classical {\it cusp} one,
we implicitly select some particular direction for the polarization of the quantum states 
relative to the direction of the emitted momentum.

By putting in eq.(\ref{main}) the classical {\it cusp} expression for $\xi\cdot I$  eq.(\ref{general}) 
one expects that, for the states corresponding to the classical {\it cusp}, it holds
$n^{\gamma+1}<a_{-n}a_{n}>=A$ with $A$ constant, strictly speaking for $n>>1$ 
(remember that we are considering just the Left component).

For the classical $cusp$  
$\sum^{n^c}_{1} n^{\gamma+1}a_{-n}a_{n}=A\cdot n^c$ divergent for $n^c\to\infty$.
\vskip0.2cm
However, the classical behaviour can only hold  up $n\leq n^c << N^{1/2}$, that is  
when the radiated energy $p_0=n/\sqrt{\alpha' N}$ is much less than the inverse 
of the string length $1/\sqrt{\alpha'}$.

Thus we assume  $n^c \sim N^{\alpha}$ with $0<\alpha <1/2$ and {\it we take as a definition 
of quantum cusp states the requirement 
$\sum^{n^c}_{1}n^{\gamma}\xi\cdot I(n)= \sum^{n^c}_{1} n^{\gamma+1}<a_{-n}a_{n}>=A\cdot n^c$.}

\vskip0.2cm
As for the value of $A$: in the literature \cite{book, Damour} it is assumed that for a generic cusp or kink 
$A$ is of the order of $N$. 
\footnote{when comparing, remember that our convention for the temporal gauge is 
$X^{0}=\alpha' M\tau$ whereas in the literature on cosmic strings it is often 
$X^{0}=\tau$; in the latter convention $A\sim N$ corresponds to $\p^2 X_T \sim 1/M$ .}
We keep this assumpion to see its implications. Actually we will find that $A\sim N$ corresponds
to  quite rare, rather than generic, configurations. 

There is a constraint on $A$ since it must be  
$\sum^{n^c}_{1}< na_{-n}a_{n}>= qN$ with $q<1$. For large $n_c$ this means that
$A=q/k(\gamma) N$ (for the standard $\gamma=4/3$ cusp $k(\gamma)\sim 3.6$). 
We take $q$ as a parameter; a very small $q$ corresponds to a rather irrelevant cusp.
\vskip0.2cm
In the interval ${\cal {I}} = \{1<<n <n^c\}$ we deform the chemical potential 
\beq
e^{-na_{-n}a_{n}\epsilon}\to e^{-n^{\gamma+1}a_{-n}a_{n}\eta} 
\eeq
while keeping $e^{-na_{-n}a_{n}\epsilon}$ for $n> n_c$. The regions $n=O(1)$  and $n=O(n_c)$
are left unspecified as they do not play an important role in the following.
  
We get the spectrum
\be
<na_{-n}a_{n}>&=& D_T{ne^{-n^{\gamma+1}\eta}\ov 1-e^{-n^{\gamma+1}\eta}}
~~for~n\subset {\cal {I}} \\
<na_{-n}a_{n}>&=& D_T{ne^{-n\epsilon}\ov 1-e^{-n\epsilon}}
~~~~~for~n>n^c
\ee

{\bf 1)} Fix $\eta$ requiring $\sum^{n^c}_{1}< n^{\gamma+1}a_{-n}a_{n}>=A\cdot n^c$.\\
By taking $n_c \leq N^{1\ov\gamma +1}$ we have
the solution  $\eta\sim D_T A^{-1}$ since in this case for $n\subset {\cal {I}}$ we
recover the {\it {cusp spectrum}}
\beq
<na_{-n}a_{n}>=D_T{ne^{-n^{\gamma+1}\eta}\ov 1-e^{-n^{\gamma+1}\eta}}\to A{1\ov n^{\gamma}}
\eeq

\vskip0.5cm

{\bf 2)} In order to fix $\epsilon$ we require
\beq
(1-q)N=D_T\{\sum_{n^c}^\infty{n e^{-n\epsilon}\ov 1-e^{-n\epsilon}}
+\sum_{1}^\infty{ne^{-n\epsilon}\ov 1+e^{-n\epsilon}}\}
\eeq
For large $n$ we approximate the sum with the integral, like in the previous section. We have
\be
\sum_{n^c}^\infty{n e^{-n\epsilon}\ov 1-e^{-n\epsilon}}\to 
{1\ov\epsilon^2}\int_{n_c\epsilon}^\infty dx x {d\ov dx}\log[1-e^{-x}] \\ \nonumber
={1\ov\epsilon^2}(-\int_{n_c\epsilon}^\infty dx \log[1-e^{-x}]-n_c\epsilon\log[1-e^{-n_c\epsilon}])
\approx -{1\ov\epsilon^2}\int_{0}^\infty dx \log[1-e^{-x}]={c_B\ov\epsilon^2}
\ee

The pre-last step holds for $n_c\epsilon\to 0$. 
In fact, for $N$ large we get $(1-q)N={c_B+c_F\ov\epsilon^2}$ and thus 
$\epsilon = N^{-1/2}\sqrt{D_T(c_F+c_B)/(1-q)}$. 
 \vskip0.5cm

{\bf 3)} The number of these {\it {cusp}} states is 
\beq
\log[{\cal N_\gamma}]=D_T\{-\sum^{n^c}_{1} \log[1-e^{-n^{\gamma+1}\eta}] 
-\sum_{n^c}^\infty \log[1-e^{-n\epsilon}]+\sum_{1}^\infty\log [1+e^{-n\epsilon}]\} 
\eeq
Note that 
\beq
-\sum_{n^c}^\infty \log[1-e^{-n\epsilon}]+\sum_{1}^\infty\log [1+e^{-n\epsilon}]\approx
{1\ov\epsilon}(c_B+c_F+n_c\epsilon\log[1-e^{-n_c\epsilon}])
\eeq

\vskip1cm

{\bf 4)} In conclusion we get 
\beq
\log[{\cal N_\gamma}]=a \cdot\bar n +2\sqrt{(1-q)D_T(c_F+c_B)}\cdot N^{1\ov 2}
-n_c\log[N^{1\ov 2}/n_c] \label{numc}
\eeq
where $a$ is some constant and $\bar n$ is the minimum between $n_c$ and $N^{1\ov\gamma +1}$.

\vskip0.2cm

For instance in the case $\gamma=4/3$ and $n_c\sim N^{1\ov\gamma +1}$ we get for the log of the ratio of the cusp number to the general average number
\beq
\log[{{\cal N}_{4/3}\ov{\cal N}}]= -2\sqrt{D_T(c_F+c_B)}\cdot(1-\sqrt{1-q}~)\cdot N^{1/2}
+(a-b\log[N])\cdot N^{3/7}
\eeq
Therefore those cusps are very rare within the variety of the generic 
string states. 
\vskip0.5cm

That fact could have been already guessed by observing how different is the $N$ dependence of 
the spectrum for $1<<n<n_c$:
it is $<na_{-n}a_{n}>\sim N/n^\gamma$ in the {\it cusp} configurations 
whereas $<na_{-n}a_{n}>\sim N^{1/2}$ for the generic string state. \\
In the generic case the dominant 
contribution to the sum rule $\sum_n <na_{-n}a_{n}>=N$ comes from $n\sim N^{1/2}$,
whereas in the {\it cusp} configurations the region $n<< N^{1/2}$  gives a substantial fraction
of the result.

Taking a higher value for $n_c$, say $n^c\sim N^{1/2}$, would give an even smaller fraction
since the main difference with the above computation would be replacing $c_B$ with 
$\tilde c_B < c_B$ in the expression of $\log[{\cal N}_\gamma]$.

\vskip0.2cm

In order to find more abundant cusp configurations we should assume $A\sim N^{1/2}$.
In this case $q\sim N^{-1/2}$  in (\ref{numc}):
we find more states but the possibly observed radiation would be more feeble and therefore
the {\it cusp-like} characterization of the signal becomes rather marginal.  
\vskip1cm

\section{ Appendix A. Temporal and LightCone Gauges}
We consider a string at rest with four-momentum $P_{\mu}=(M,\vec 0)$.
Here we put $\alpha'=1$.

The Virasoro constraints are, for the Left or Right part,
\beq
(\p X^0_{L,R})^2=(\p Z_{L,R})^2+(\p \vec X_{L,R})^2
\eeq
where $\p$ is the derivative with respect to the World-Sheet (WS) parameter 
$s_{L,R}=\tau\pm\sigma$ which is different for different gauge choices.
We have chosen a $Z-$direction thus $\vec X$ is defined to be transverse. 
Let us consider for instance the Left part (dropping the suffix $"L"$). 
\vskip0.2cm

In the {\it Temporal (TP) gauge} one takes $\p_{\hat s} X_{TP}^0= {M\ov 2} $
where we call $\hat s$ the WS parameter.
\vskip0.2cm

In the {\it LightCone (LC) gauge} one
takes $\p_{s} X_{LC}^+ \equiv{\p_{s} X_{LC}^0+\p_{s} Z_{LC}}={M\ov 2}$ where we call $s$ the new WS parameter.
\vskip0.2cm

Classically, the passage between TP and LC is a redefinition of the WS 
parameter $\hat s\to s$, that is $X_{LC}(s)=X_{TP}(\hat s(s))$ and similarly for $X^0,Z$:
\beq
(\p_{\hat s} X_{TP}^0+\p_{\hat s}Z_{TP}) {\p\hat s\ov\p s}={M\ov 2} ~~ 
\Rightarrow ~~ {\p\hat s\ov\p s}={1\ov 1+ {2\p_{\hat s}Z_{TP}\ov M}}
\label{LC}
\eeq
Note that, because of the constraint, $|\p_{\hat s}Z_{TP}|\leq M/2$ and therefore both $\hat s(s)$ 
and $s(\hat s)$ are well defined.
It follows that
\beq
\p_s \vec X_{LC}={\p_{\hat s}\vec X_{TP}\ov 1+ {2\p_{\hat s}Z){TP}\ov M}}
\eeq
For instance, we see that even if $~\p_{\hat s}\vec X_{TP}, ~~ \p_{\hat s}Z_{TP}~$  
only contain one Fourier mode of the WS parameter in the TP (like it is for the maximal angular momentum string configuration), 
$~\p_s \vec X_{LC}~$ will in general contain all the Fourier modes of the WS parameter in the LC.
 
\vskip0.2cm

{\it Viceversa} in the LC gauge we have 
\be
 \p_s X_{LC}^-={2|\p_s \vec X_{LC}|^2\ov M} ~~ &\Rightarrow& ~~ \
\p_s X_{LC}^0={M\ov 4}+{|\p_s \vec X_{LC}|^2\ov M} \\  \label{X0}
{\p s\ov\p\hat s}\p_s X_{LC}^0={M\ov 2}~~& \Rightarrow& ~~
{\p s\ov\p\hat s}={2\ov 1+{4|\p_s \vec X_{LC}|^2\ov M^2}}
\label{uffa}
\ee
and therefore
\beq
\p_{\hat s}\vec X_{TP}={2\p_s \vec X_{LC}\ov 1+{4|\p_s \vec X_{LC}|^2\ov M^2}}
\label{XTP}
\eeq

It can be checked that (\ref{LC}) and (\ref{XTP}) are consistent.

We see that the classical relation between TP and LC is highly nonlinear.
The quantum version of it is to our knowledge not available.
\vskip1cm

\section{ Appendix B. The shape of a generic string}
From the form of the (bosonic) spectrum eq.(\ref{spectrum}) and the main formula 
eq.(\ref{main}) we can reconstruct the corresponding classical transverse string
in the LC gauge (consider here the Left component), by putting for the transverse part:
\beq
X^i_{LC}(s)=\sum_n c^i_n Cos [ns+\theta^i_n]
\eeq

where $\theta^i_n$ are random phase shifts, in general different for Left and Right, and 
\beq
\sum_i (c^i_n)^2= A{1\ov n}{e^{-{n\ov g\sqrt N}}\ov 1-e^{-{n\ov g\sqrt N}}}
\eeq

Here $g=1/\sqrt{D_T(c_B+c_F)}$ and $A=\sum A^i$ where the values of $A^i$ 
are randomly distributed among the transverse directions.

For the sake of simplicity we will consider the particular example where only one of the $X^i_{LC}$ is different from zero. Therefore our sample string is less random than the true generic one. We conventianally take $A=g=1$.

We do the same for the Right component, with different random phase shifts.

We have taken $N=100$ (and cutoff the sum at $n=100$).
We assume that the string state is at rest and its mass is, according to the LC formulae, 
\beq
M=P_+=P_-=2 \sqrt{\int_0^{2\pi} ds (\p_s X_{LC}(s))^2\ov 2\pi}
\eeq
The result is the same for Left and Right as it should be,
and $M$ is proportional to $\sqrt N$. 

According to the LC prescription we put (Left component)
\beq
X_{LC}^+ = {M\ov 2}s ~~~~
X_{LC}^- = {M\ov 2}s+\{ {\int_0^{s} ds' (\p_s X_{LC}(s'))^2\ov 2\pi}-{M\ov 2}s \}
\eeq
The part of $X^-$ which is in curly brackets is taken to be $2\pi$-periodic,
that is, its value at $s+2\pi$ is identified with its value at $s$.
Further, from the previous formulae,
\beq
Z_{LC}=- \{ {\int_0^s ds' (\p_s X_T(s'))^2\ov 2\pi}-{M\ov 2}s \}
\eeq

In order to get the string in the Temporal gauge we use eq.(\ref{uffa}) to get $s(\hat s)$ and take  
$ X_{TP}(\hat s)=X_{LC}(s(\hat s)),~  Z_{TP}(\hat s)=Z_{LC}(s(\hat s))$.
 
We do the same for the Right part and finally we get in the TP gauge 
\be
X_{TP}(\tau ,\sigma)&=& X_{TP}(Left)(\tau-\sigma)+X_{TP}(Rigth)(\tau+\sigma) \\ \nonumber
Z_{TP}(\tau ,\sigma)&=&Z(_{TP}Left)(\tau-\sigma)+Z_{TP}(Rigth)(\tau+\sigma)
\ee

In Fig.1,2,3) we show the resulting (TP) string in the plane $X,Z$ for some values of  \\
$\tau=0,\pi/4,\pi/2$.
\vskip0.5cm
\begin{figure}[ht]
\centerline{
\epsfig{file=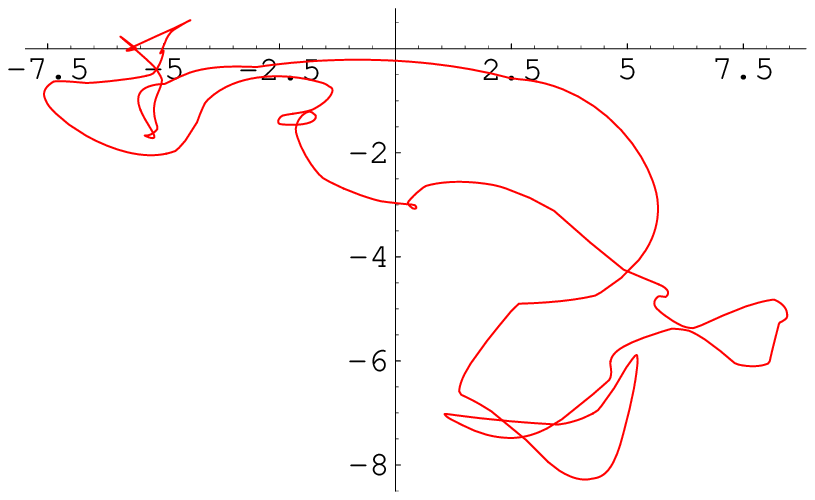,width=.4\textwidth}
Fig.1
\qquad
\epsfig{file=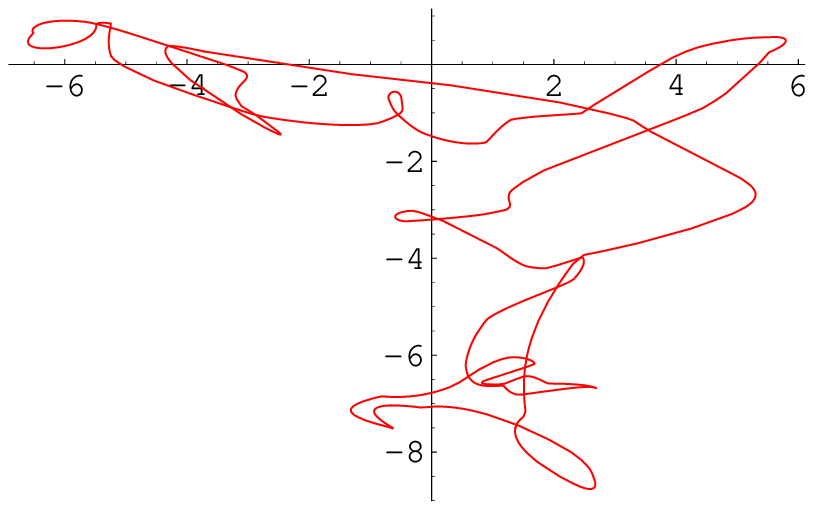,width=.4\textwidth}
Fig.2}
\vskip0.5cm
\centerline{
\epsfig{file=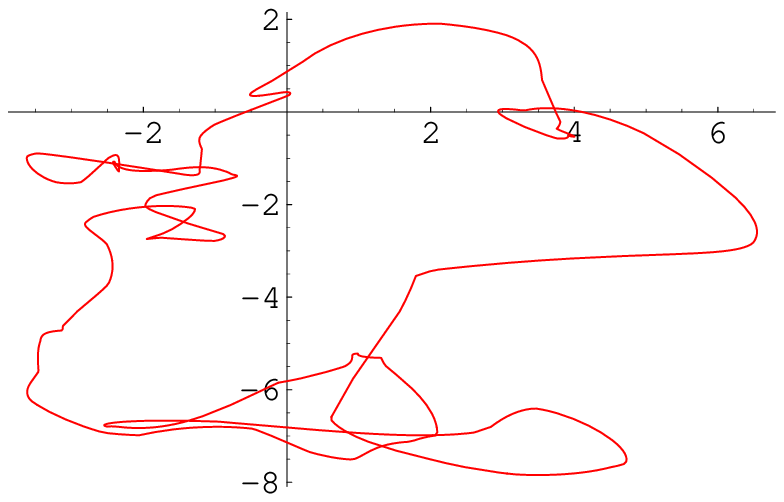,width=.4\textwidth}
Fig.3}
\end{figure}

\newpage

\section{ Appendix C. A particular cusp-like case: the maximal angular momentum state}

The classical state of a closed string of maximal angular momentum is, in the Temporal Gauge,
\beq
{X_L^1+iX^2_L\ov\sqrt{2}}= {L\ov 2\sqrt{2}} e^{i(\tau+\sigma)} ~~~~~{X_R^1+iX^2_R\ov\sqrt{2}}= 
{L\ov 2\sqrt{2}} e^{i(\tau-\sigma)} 
\eeq
that is 
\beq
{X^1+iX^2\ov\sqrt{2}}={X_L^1+iX^2_L\ov\sqrt{2}}+ {X_R^1+iX^2_R\ov\sqrt{2}}=
{L\ov\sqrt{2}}e^{i\tau}cos(\sigma)
\eeq
and $X^0=\alpha' M\tau =L\tau$. In the quantum state $M=2\sqrt{N/\alpha'}$ where the integer 
$N$ is the eigenvalue of the number operator $\hat N$.

One would think to represent the corresponding quantum state by a coherent state of the oscillators.
However due to the Virasoro constraints this coherent state would not have a definite mass and therefore
$X^0$ would be undefined. Luckily we do not need that, since we know precisely 
the unique quantum maximal angular momentum state in the Temporal Gauge:
\beq
|\Psi^{Jmax}>={(b_{-1})^N\psi^b_{-1/2}[0>_L\ov \sqrt{N!}}
\otimes{(\tilde b_{-1})^N\tilde\psi^b_{-1/2}[0>_R\ov \sqrt{N!}}
\eeq
where $b_{-1}={a^1_{-1}+ia^2_{-1}\ov\sqrt{2}}, ~\psi^b_{-1/2}={\psi^1_{-1/2}+i\psi^2_{-1/2}\ov\sqrt{2}}$.
Therefore we can compute $\sum_{\xi} |\xi\cdot I_{L}|^2$ both classically and quantum mechanically.

Classically one has radiation of the bosonic part of the graviton multiplet therefore we compare with
the quantum  NS massless emission. 

The relevant formulae are written in \cite{CIR2}, that is, referring to \cite{CIR2},
the modulus square of eq.(3.45) for the classical radiation and eqs.(3.7,8) for the quantum computation, together with the explicit expressions in (3.10,11,12,13) and in Appendix.B for the rest. It is important to keep
all the terms of the quantum computation, which has been checked by comparing the result for
$NS_L\times NS_R$ with 
the independent computation made by taking the imaginary part of the torus diagram and restricting the spinstructure to NS-NS.

To look for the {\it cusp} spectrum we take the emitted momentum to lay in the $X^1,X^2$ plane.
The classical result for $ \sum_{\xi}|\xi\cdot I_{L}|^2$  is expressed as a constant times $N$ times
a function depending on $n$ only (remember the emitted energy $\omega =n/(2\sqrt{N})$).
It  reaches rather slowly the expected behavior  $n^{-4/3}$ for large $n$.

The quantum result is a more complicated non factorized expression in terms of $N$ and $n$
and we have computed it for $N=1000$ and $n\leq 300$.
It is very near to the classical result for $n< 50$ (this is also a check of the computation since 
the normalization is fixed), where however the classical result has not yet reached the {\it cusp-like}
asymptotic behaviour, after which it goes to zero more rapidly.

The comparison is shown in Fig.4 where we show the classical (black) and quantum (red) results as a function of $n$,
both multiplied by $c\times n^{4/3}$, choosing $c$ such as to get the classical curve $=1$ for $n=1000$.
\begin{figure}[ht]
\centerline{
\epsfig{file=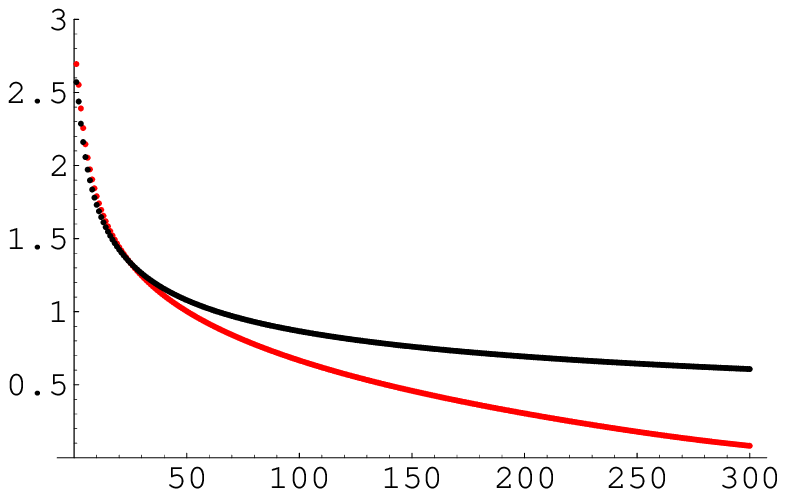,width=.4\textwidth}}
\centerline{Fig.4}
\end{figure}

Therefore this particular state does not seem to follow the {\it cusp-like} pattern and therefore its behaviour cannot be compared with the average {\it cusp} one.  However it agrees with the generic expectation that the most important part of the radiation is emitted for low $n$ where it matches the classical pattern.
 In general, that region of small $n$ is not likely to be part of the asymptotic, possibly {\it cusp-like}, behaviour. If this is true, then the {\it cusp} characterization of the string states would not be so relevant for observations.

\bigskip

\section{Acknowledgements}
The author is indebted with J.Russo for reading the manuscript and making useful comments.
The author acknowledges partial support by the EC-RTN network
MRTN-CT-2004-005104.

\end{document}